\documentclass[lettersize,journal]{IEEEtran}
\usepackage{enumitem}
\usepackage{amsmath}
\usepackage{graphicx}
\usepackage{amssymb}
\usepackage[colorlinks=true, allcolors=blue]{hyperref}
\usepackage{listings}
\usepackage[table]{xcolor}

\def\tablename{Table}

\definecolor{LightCyan}{rgb}{0.88,1,1}

\colorlet{punct}{red!60!black}
\definecolor{background}{HTML}{EEEEEE}
\definecolor{delim}{RGB}{20,105,176}
\colorlet{numb}{black}
\lstdefinelanguage{json}{
	basicstyle=\footnotesize\ttfamily,
	numberstyle=\scriptsize,
	stepnumber=1,
	numbersep=8pt,
	showstringspaces=false,
	breaklines=true,
	frame=lines,
	literate=
	*{0}{{{\color{numb}0}}}{1}
	{1}{{{\color{numb}1}}}{1}
	{2}{{{\color{numb}2}}}{1}
	{3}{{{\color{numb}3}}}{1}
	{4}{{{\color{numb}4}}}{1}
	{5}{{{\color{numb}5}}}{1}
	{6}{{{\color{numb}6}}}{1}
	{7}{{{\color{numb}7}}}{1}
	{8}{{{\color{numb}8}}}{1}
	{9}{{{\color{numb}9}}}{1}
	{:}{{{\color{punct}{:}}}}{1}
	{,}{{{\color{punct}{,}}}}{1}
	{\{}{{{\color{delim}{\{}}}}{1}
	{\}}{{{\color{delim}{\}}}}}{1}
	{[}{{{\color{delim}{[}}}}{1}
	{]}{{{\color{delim}{]}}}}{1},
}

\begin{document}
\title{NetSatBench: A Distributed LEO Constellation Emulator with an SRv6 Case Study}
\author{Andrea Detti, Shahram Dadras, Giuseppe Tropea \thanks{Andrea Detti and Shahram Dadras are with CNIT and the Department of Electronic Engineering, University of Rome "Tor Vergata", Italy. Giuseppe Tropea is with NetSense srl. The study is partially funded by GeminiPort, proj. n. FTE0000471 CUP B67H22005960008}}

\maketitle

\begin{abstract}
NetSatBench is a distributed emulation platform for evaluating communication protocols and application workloads over large-scale LEO satellite systems. Satellites, gateways, and user terminals are implemented as Linux containers distributed across a cluster of bare-metal or virtual machines, while emulated links are realized through a Layer-2 VXLAN overlay. The system state is maintained in an Etcd key-value store and updated through epoch files, which propagate link and task changes to local control agents running inside the emulated nodes. In contrast to library-oriented tools that require users to write control programs, NetSatBench adopts a higher-level declarative workflow based on JSON "scenario files" and a command-line interface. The platform decouples physical-layer and routing modeling from the emulator core through external plug-ins, while providing built-in support for IPv4 and IPv6 routing, including IS-IS and ideal time-varying routing. Rather than focusing on emulator micro-performance alone, we illustrate what NetSatBench enables through an SRv6-based LEO architecture in which control procedures manage data tunnels between users and gateways under different handover policies. This case study shows how NetSatBench can support protocol-level experimentation under time-varying LEO dynamics and highlights the importance of end-to-end handover strategies that jointly account for the satellites serving both the user and the gateway.
   
\end{abstract}

\section{Introduction}
\label{sec:intro}
Low Earth Orbit (LEO) satellite networks are rapidly evolving from a long-standing vision into a practical networking platform for global Internet access and wide-area distributed services. This transition is driven by recent advances in satellite technology, lower launch costs enabled by reusable rockets, and the deployment of large constellations such as SpaceX Starlink and Eutelsat OneWeb. As a result, the systems community now faces a concrete engineering problem: how to design, validate, and compare network mechanisms for large, dynamic constellations before they are deployed in orbit.

Modern LEO constellations provide performance characteristics that make this problem especially relevant. Their relatively low orbital altitude, typically between 500 and 2000 km, enables end-to-end round-trip times on the order of 20--70 ms, substantially below those of geostationary systems. In parallel, multi-antenna satellite-to-ground communication and beam-based access support throughput in the hundreds of Mbps \cite{mohan2024multifaceted}, and the progressive adoption of optical inter-satellite links (ISLs), offering capacities of several tens of Gbps, is further transforming these systems from bent-pipe access networks into multi-hop space networks. Consequently, LEO constellations are becoming a realistic substrate for global Internet service providers. 

However, designing protocols and services for this environment is difficult because LEO networks combine large scale, topology dynamics, and limited tolerance for post-deployment corrections. Once satellites are launched, fixing software defects or revising protocol behavior becomes significantly more difficult and costly than in terrestrial systems \cite{lai2025leocc,cao2023satcp, spacecloud2021}. For this reason, realistic pre-deployment experimentation is essential with emulation platforms that can execute real software stacks while reproducing time-varying connectivity, delays, and handover behavior at constellation scale.

Existing tools only partially satisfy this need. Single-host emulation platforms simplify deployment, but their architecture inherently limits scalability and constrains the complexity of the experiments that can be executed \cite{milla2026emerging,cao2023satcp}. Recent distributed systems, including OpenSN \cite{lu2025opensn} and StarryNet \cite{lai2023starrynet}, improve the scalability of LEO-network emulation by distributing the emulation workload across a cluster of worker nodes and by exposing \textit{library-based} interfaces for defining scenarios and experiment behavior through Python programs.

This paper presents \emph{NetSatBench}, a distributed emulation platform that complements systems such as StarryNet and OpenSN through a cleaner separation between the emulation core and experiment-specific plug-ins, such as routing and physical-layer models, together with a higher-level \textit{declarative} workflow for defining scenarios and experiment behavior through JSON configuration files. NetSatBench distributes emulated nodes across a cluster of worker nodes coordinated through Etcd and container-resident agents, realizes links through VXLAN tunnels, and drives time-varying behavior through declarative epoch files produced either manually or by an extensible physical-modeling pipeline. Besides describing the network state at each instant, epoch files can also encode scheduled tasks, thereby supporting experiments with time-triggered actions.

The evaluation of new routing mechanisms and physical-layer models is supported by lightweight Python interfaces that integrate external models as plug-ins without requiring modifications to the emulator core. On the routing side, NetSatBench provides built-in plug-ins for IS-IS and local-link IPv4/IPv6 routing, as well as ideal precomputed min-hop and min-delay routing, which can serve either as benchmarks or as a way to abstract away routing when the experiment focuses on other aspects. The platform also supports configurable automatic IPv4/IPv6 addressing and name resolution. On the physical-layer side, the same plug-in approach enables custom models for latency, bitrate, packet loss, and link visibility. Finally, a command-line interface with \texttt{key:value} property matching simplifies software deployment and execution on selected nodes, as well as the collection of experiment results and node configurations at scale.

\begin{figure}
    \centering
    \includegraphics[width=0.8\linewidth]{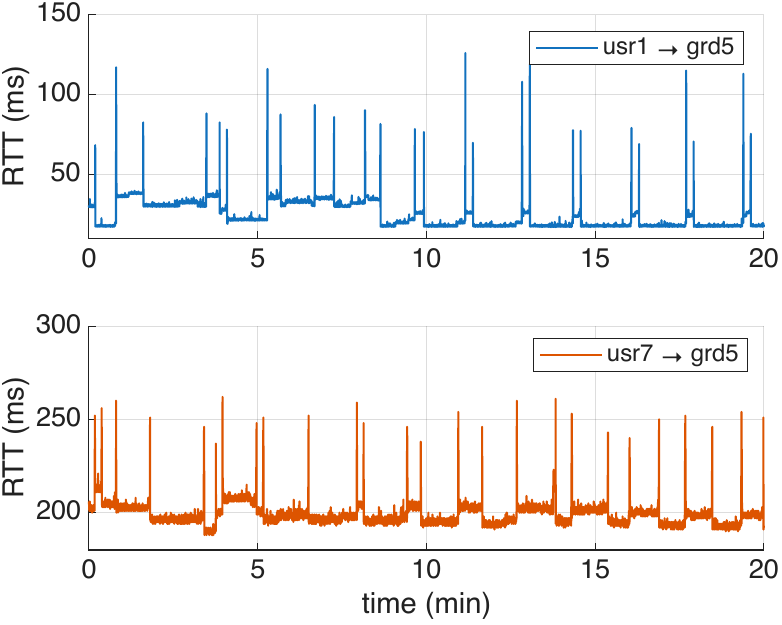}
    \caption{RTT measurements between two users and the gateway in a OneWeb-like scenario. Local min-access-delay handover strategy.}
    \label{fig:intro-rtts}
\end{figure}

To illustrate this capability, we use NetSatBench to study an IPv6 Segment Routing (SRv6)-based LEO architecture that realizes \textit{satellite PDU sessions}, i.e., data tunnels between users and their ground gateways. In this design, time-dependent precomputed IPv6 forwarding tables are installed within the infrastructure domain, while dynamic SRv6 tunnels connect gateways and user terminals in a manner analogous to 5G PDU sessions. The case study considers a OneWeb-like constellation with 588 satellites, different handover strategies, and TCP congestion-control algorithms including Cubic and BBR. Rather than emphasizing emulator micro-performance, we use this case study to show the kind of experimentation that the platform enables. To support practical understanding, reproducibility, and adoption, we release NetSatBench and the complete use-case artifacts as open-source software, including the configuration files and command-line workflow used for the experiments \cite{netsatbench}. \autoref{fig:intro-rtts} previews this capability by showing RTT measurements between two user terminals and their gateways, illustrating how NetSatBench can reproduce realistic LEO-network behavior, including abrupt delay variations and transient spikes caused by handover events, consistent with RTT patterns reported for operational LEO constellations \cite{mohan2024multifaceted,huston2024starlink}.

Overall, the main contribution of this paper is twofold. First, we design and implement \emph{NetSatBench}, a distributed platform for evaluating real network protocols and applications in large LEO constellations. Second, we devise an SRv6-based LEO architecture as a representative use case to demonstrate how NetSatBench can evaluate end-to-end solutions under realistic operating conditions and can surface protocol-level insights, such as the importance of end-to-end handover decisions.

The remainder of the paper is organized as follows. \autoref{sec:design} presents the design of NetSatBench. \autoref{sec:usecase} describes the SRv6 use case. \autoref{sec:evaluation} reports the experimental results. \autoref{sec:related} discusses related work, and \autoref{sec:conclusion} concludes the paper.

\section{NetSatBench}
\label{sec:design}

\subsection{Design Goals and Scope}

The NetSatBench design is guided by five goals:

\begin{itemize}
\item \textbf{Scalability:} Supports constellations with hundreds to thousands of nodes (satellites, gateways, users) through a distributed, container-based architecture coordinated by an efficient publish-subscribe control plane.

\item \textbf{Multi-layer experimentation:} Enables integrated evaluation across layers, from physical/link models to routing, transport, and applications, including cross-layer interactions and in-orbit distributed services. The SRv6 use case (\autoref{sec:usecase}) demonstrates this capability by jointly exploring link models, routing, and end-to-end tunnel reconfiguration.

\item \textbf{Extensibility:} Provides modular interfaces for custom physical-layer and routing models via lightweight Python plug-ins, allowing rapid prototyping of new mechanisms without modifying the emulation code.

\item \textbf{Reproducibility:} Uses JSON-based declarative configurations and epoch-based descriptions to fully specify system state and dynamics, enabling controlled comparisons of alternative designs under identical conditions.

\item \textbf{Operational support:} Integrates topology configuration, deployment, execution, monitoring, and data upload and download through a unified Python CLI, provided by the \texttt{nsb} command, that supports \texttt{key:value} matching reducing the operational complexity of large-scale experiments in clusters with many workers and emulated nodes.
\end{itemize}



\subsection{System Architecture}

\begin{figure*}
    \centering
    \includegraphics[scale=0.8]{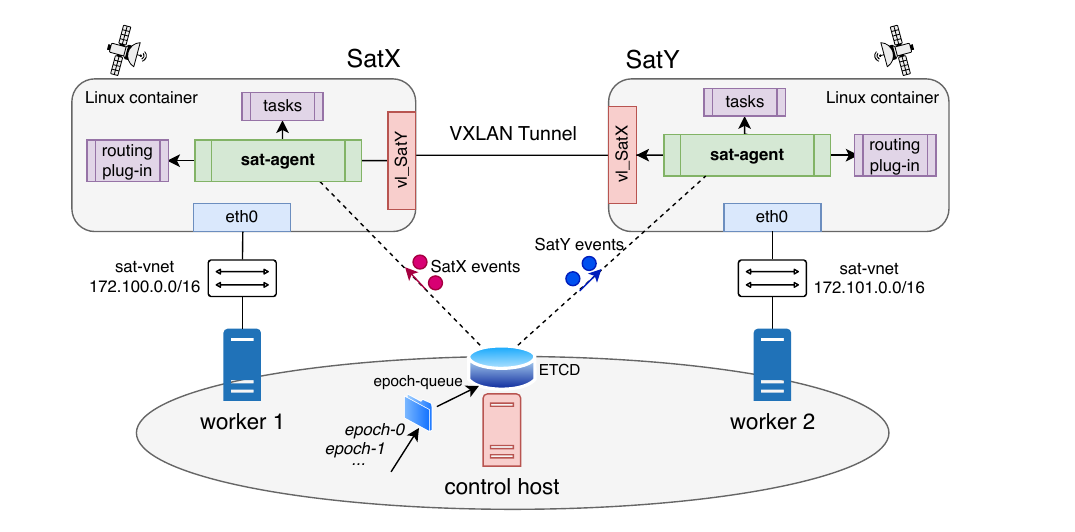}
    \caption{Architecture of the NetSatBench emulator.}
    \label{fig:arch}
\end{figure*}

\autoref{fig:arch} illustrates the architecture of NetSatBench. The platform runs on a set of Linux hosts, called \textit{workers}, interconnected through an Ethernet network referred to as the \textit{underlay}. Each worker hosts multiple emulated nodes as Docker containers, representing satellites, gateways, and user terminals, hereafter collectively referred to as \textit{nodes}. To avoid confusion, in the following we refer to the person using NetSatBench as the \textit{researcher}, while the term user refers to the emulated user terminal. 

Nodes are interconnected through a Layer~2 \textit{overlay} network implemented via VXLAN tunnels, which realize the links of the emulated system, including inter-satellite links (ISLs), satellite–gateway links (SGLs), and satellite–user links (SULs). Link characteristics such as bitrate, delay, and packet loss are enforced using the Linux Traffic Control framework.

Each node container connects to the underlay through a local \texttt{eth0} interface attached to a common Linux bridge on its hosting worker, and VXLAN endpoints are bound to these interfaces, thereby terminating directly in the node containers. The \texttt{eth0} interfaces are assigned addresses from a private IPv4 subnet, denoted \texttt{sat-vnet}, that is unique per worker (e.g., 172.100.0.0/16 for worker-1). Workers are configured to route these subnets across the cluster, ensuring full connectivity among node containers and enabling the establishment of VXLAN tunnels.

The nodes can also reach external networks through their hosting worker, enabling hybrid experimentation in which the emulated gateway nodes provide real Internet connectivity to the user terminals. This capability allows for the evaluation of LEO satellite segments integrated with the operational Internet.

The control plane consists of three components:
\begin{itemize}
  \item A distributed key-value store based on Etcd\footnote{\url{https://etcd.io}.}, which maintains the current state of the emulation.
  \item Local control agents, called \textit{sat-agents}, which run in each node's container. Each \textit{sat-agent} subscribes to the relevant Etcd entries and reacts to state changes by configuring its container, for example, by creating VXLAN interfaces, applying traffic control policies, and executing runtime commands.
  \item A control host with the \texttt{nsb} command-line interface (CLI), which is used to define and manage the emulation scenario, deploy it across workers, control its evolution, transfer software to nodes, and collect experimental data.
\end{itemize}

\subsection{Emulation State: Representation and Update}
\label{sec:state}

NetSatBench organizes execution into discrete time intervals called \textit{epochs}, each defined by a JSON file. The system state is updated only at the epoch boundaries and is maintained in the Etcd key-value store, which serves as the single source of truth for the global emulation state.

\autoref{l:etcd-keyspace-compact} shows representative Etcd entries, organized into static entries and epoch-driven entries. Static entries capture persistent properties of workers (under \texttt{/config/workers/}) and nodes (under \texttt{/config/nodes/}). Worker entries record the underlay IP address, available resources, and SSH credentials. The node entries specify the node type (satellite, gateway, or user), resource requirements and limits, worker placement, the container's \texttt{eth0} address, and optional Layer~3 configuration such as routing modules (plug-in) and related metadata.

For experiments requiring Layer~3 functionality, each node is assigned a dedicated IPv4 and/or IPv6 subnet in CIDR notation, either manually or by automatic address assignment. The last address in the subnet is bound to the node's loopback interface and used for communication over the emulated links. Name resolution is handled through mappings between node names and loopback addresses, which the sat-agent injects into each container's \texttt{/etc/hosts} file so that applications can refer to other nodes by name.

Epoch-driven entries describe the time-varying part of the emulation. Link entries (under \texttt{/config/links/}) define per-link parameters such as bitrate, delay, and packet loss, while task entries (under \texttt{/config/run/}) specify commands to be executed on selected nodes at the beginning of an epoch.

State updates are driven by epoch files, as illustrated in \autoref{l:epoch-file-example}. Each epoch file contains the epoch start time together with lists of link events (additions, updates, and deletions) and runtime tasks to be executed on specific nodes.
The control tool \texttt{nsb run} processes these epoch files and updates Etcd accordingly. It supports two execution modes. In \textit{real-time mode}, epoch files are placed into a dedicated \texttt{epoch-queue} folder and processed immediately, allowing external tools such as a digital twin of the satellite system to inject topology changes as they occur. In \textit{discrete-time mode}, \texttt{nsb run} sequentially reads pre-generated files (epoch-0, epoch-1, \ldots) from a user-specified directory and pushes them into the \texttt{epoch-queue} while preserving their timing relationships, thereby enabling fully repeatable experiments.

Etcd propagates epoch-file state updates to the sat-agents through a publish-subscribe mechanism. The structure of the Etcd key space (\autoref{l:etcd-keyspace-compact}) allows each sat-agent to subscribe only to the changes relevant to its node, such as link-parameter updates (e.g., \texttt{/config/links/sat1/}) and scheduled tasks (e.g., \texttt{/config/run/sat1}). This design is efficient because it confines update transmission and processing to the nodes that must react to each change.

\begin{figure}[t]
  \centering
  \begin{lstlisting}[language=json,basicstyle=\scriptsize\ttfamily,caption={Representative Etcd key-value entries describing the emulation state},captionpos=b,label={l:etcd-keyspace-compact},escapeinside=||]
|\color{red}\textit{--- static configuration entries ---}|
|\color{gray}\textit{workers configuration}|
 "/config/workers/worker-1":{
  "ip": "10.0.1.144", "cpu": "4", "mem": "8GiB",
  "sat-vnet-cidr": "172.100.0.0/16",
  "ssh-user": "ubuntu", "ssh-key": "~/.ssh/id_rsa"}
 ...
|\color{gray}\textit{nodes full-configuration}|
 "/config/nodes/sat1":{
  "type": "satellite",
  "image": "msvcbench/sat-container:latest",
  "cpu-request": 0.1, "mem-request": "100MiB",
  "cpu-limit": 1, "mem-limit": "200MiB",
  "worker": "worker-1",
  "eth0_ip": "172.100.3.2",
  "L3-config": {
   "enable-routing": true,
   "routing-module": "extra.routing.isisv6",
   "cidr-v6": "2001:db8:100::/126"}}
  ...
|\color{gray}\textit{node name to IP mapping}|
 "/config/etchosts6/sat1": "2001:db8:100::3"
...
|\color{red}\textit{--- epoch-driven configuration entries ---}|
|\color{gray}\textit{current epoch info}|
 "/config/epoch-config":{
  "epoch-time": "2023-10-01T00:22:41Z",
  "epoch-file": "NetSatBench-epoch600.json",...}
|\color{gray}\textit{links configuration}|
 "/config/links/sat1/vl_sat2":{
  "endpoint1":"sat1","endpoint2":"sat2",
  "rate":"400mbit","loss":0,"delay":"3ms"}
...
|\color{gray}\textit{tasks to execute}|
 "/config/run/sat1": ["tcpdump -i vl_sat2"]
...
\end{lstlisting}
\end{figure}

\begin{figure}[t]
 \centering
 \begin{lstlisting}[language=json,basicstyle=\scriptsize\ttfamily,caption={Representative JSON epoch file format},captionpos=b,label={l:epoch-file-example}, escapeinside=||]
|\color{gray}\textit{epoch time}|
"time": "2023-10-01T00:22:41Z",
|\color{gray}\textit{link delete}|
"links-del": [
 {"endpoint1": "sat1", "endpoint2": "grd1"}],
|\color{gray}\textit{link parameter update}|
"links-update": [
 {"endpoint1": "sat1", "endpoint2": "sat2", "rate": "400.0mbit", "loss": "0.0", "delay": "3.0ms"}],
|\color{gray}\textit{link addition}|
"links-add": [
 {"endpoint1": "sat1", "endpoint2": "usr1", "rate": "23.0mbit", "loss": "0.0", "delay": "6.0ms"}],
|\color{gray}\textit{tasks to execute}|
"run": {
 "sat1": ["ip route replace ..."]}
  \end{lstlisting}
\end{figure}

\begin{figure*}
    \centering
    \includegraphics[width=0.8\linewidth]{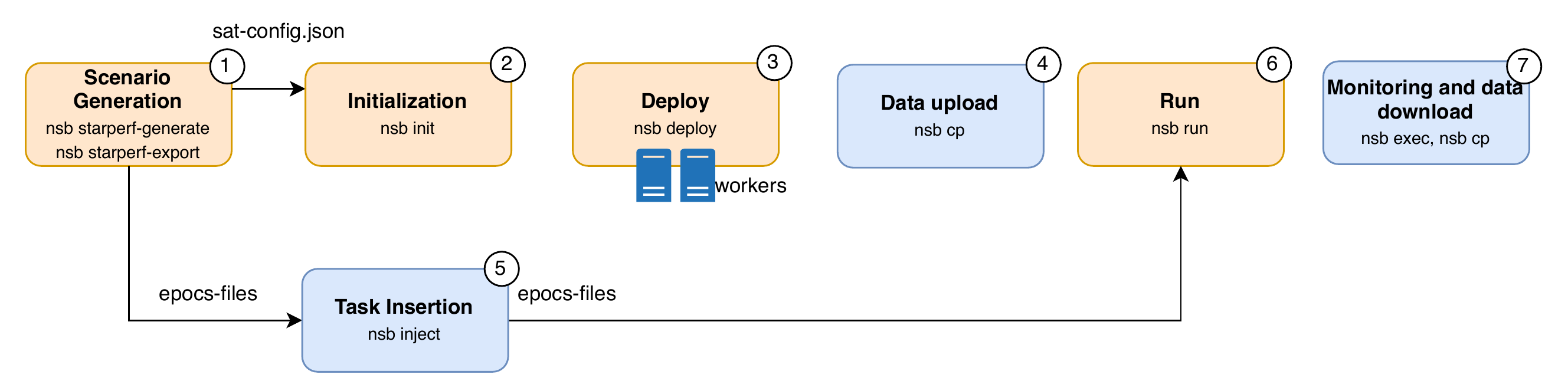}
    \caption{Experiment workflow}
    \label{fig:control-pipeline}
\end{figure*}

\begin{figure}[t]
  \centering
  \begin{lstlisting}[language=json,basicstyle=\scriptsize\ttfamily,caption={Example NetSatBench configuration for node defaults and epoch settings},captionpos=b,label={l:sat-config}, escapeinside=||]
|\color{gray}\textit{common node configuration and IP auto-assignment}|
"node-config-common": [
 {"match-key": "type", "match-value": "satellite",
  "config-common": {
   "image": "msvcbench/sat-container:latest",
   "cpu-request": "100m",
   "mem-request": "100MiB",
   "L3-config": {
    "enable-routing": true,
    "routing-module": "extra.routing.isisv6",
    "auto-assign-ips": true,
    "auto-assign-super-cidr": [{
     "match-key": "type","match-value": "satellite",
     "super-cidr6": "2001:db8:100::/48"}]}}},...],
|\color{gray}\textit{specific node configuration}|
"nodes": {
 "sat1": {"type": "satellite"},
 "grd1": {"type": "gateway"},
 "usr1": {"type": "user"},...}
|\color{gray}\textit{epoch configuration}|
"epoch-config": {
 "epoch-dir": "examples/StarPerf/OneWeb/epochs",
 "file-pattern": "NetSatBench-epoch*.json"},
  \end{lstlisting}
\end{figure}

\subsection{Experiment Workflow}
\label{sec:workflow}

\autoref{fig:control-pipeline} illustrates the end-to-end workflow of an experiment. The figure also reports the command-line instructions used to execute each step of the workflow (e.g., \texttt{nsb init} and \texttt{nsb deploy}). For brevity, the figure omits the initial software and network configuration of the worker hosts, which is nonetheless fully automated through the \texttt{nsb} command line.

\subsubsection*{Scenario Generation}
The first step is to generate the emulated scenario, which consists of a \texttt{sat-config.json} file containing a \textit{minimal} definition of the emulated nodes and a sequence of JSON epoch files. As shown in \autoref{l:sat-config}, the \texttt{sat-config.json} file specifies both common properties and per-node configurations. Common properties are applied to nodes through a \texttt{key:value} matching rule, where the key and value are defined in each node's specific configuration. For example, the common configuration shown in \autoref{l:sat-config} applies to all nodes whose specific configuration includes \texttt{type: satellite}, such as \texttt{sat1}.    

For small-scale or educational experiments, configuration and epoch files can be authored manually. For larger deployments, they are typically generated automatically by a physical-layer pipeline that accepts researcher-defined parameters describing the constellation, ground infrastructure, and physical-layer models. Further details on the built-in physical-modeling pipeline are provided in \autoref{sec:physical-modeling}.

\subsubsection*{Initialization}
During the initialization phase, NetSatBench reads the node configuration and epoch files and writes to Etcd the complete configuration of each node, after merging the matching common properties, as shown in \autoref{l:etcd-keyspace-compact}. During the same phase, it assigns IPv4/IPv6 subnets and determines the worker on which each node will run.

Worker scheduling is both topology-aware and resource-aware. The scheduler analyzes all epoch files and assigns each emulated link a weight equal to the number of epochs in which that link is active. It then applies an iterative METIS-based graph-partitioning procedure~\cite{karypis1998multilevel}, starting from a single partition and progressively increasing the number of partitions until the cumulative requested resources of each partition fit within the capacity of a single worker\footnote{Similarly to Kubernetes, resource request values are used for scheduling, while resource limit values define hard caps enforced by the container engine.}. This approach co-locates frequently communicating nodes on the same worker whenever possible, thereby reducing traffic contention on the underlay network.

\subsubsection*{Deploy}
During the deployment phase, NetSatBench reads the worker and node configuration from Etcd and instantiates the corresponding containers on their assigned workers by contacting the remote Docker engines via SSH. When a container is created, the local sat-agent publishes its \texttt{eth0} address to Etcd for use as the VXLAN endpoint, configures the node's loopback IP address, and updates the name-resolution entries in Etcd. The sat-agent then starts monitoring the node-specific link and task entries in Etcd to track subsequent topology changes during the run phase.

\subsubsection*{Run and Optional Utilities}
After deployment, NetSatBench processes epoch events and updates the system state in Etcd, which propagates the corresponding changes to the relevant sat-agents, as described in \autoref{sec:state}. Optional utilities support experiment automation and data collection. If epoch files do not already contain task commands in the \texttt{run} section, users can add them manually or with \texttt{nsb inject}. Software can be transferred to and from emulated nodes with \texttt{nsb cp}, while \texttt{nsb exec} runs commands on selected nodes, for example to open a shell or modify a network parameter. These utilities support node selection either by explicit name or by \texttt{key:value} matching, making it easy to target classes of nodes such as all gateways or all users.

\subsection{Built-in Physical-Modeling Pipeline}
\label{sec:physical-modeling}

\begin{figure}
    \centering
    \includegraphics[width=1.0\linewidth]{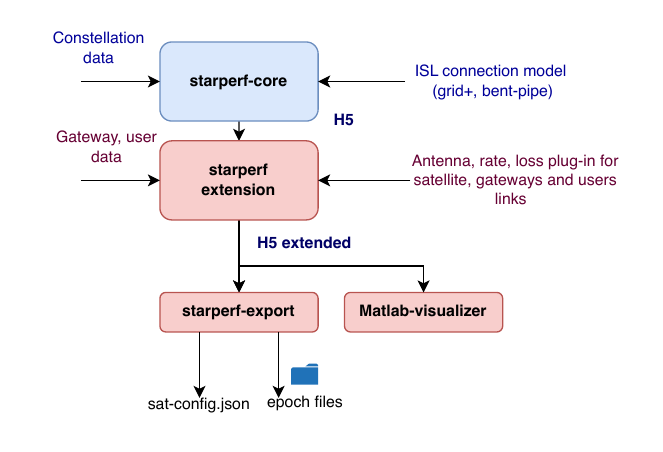}
    \caption{Scenario-generation pipeline based on StarPerf and NetSatBench extensions.}
    \label{fig:generator-pipeline}
\end{figure}

\begin{figure}
    \centering
    \includegraphics[width=0.9\linewidth]{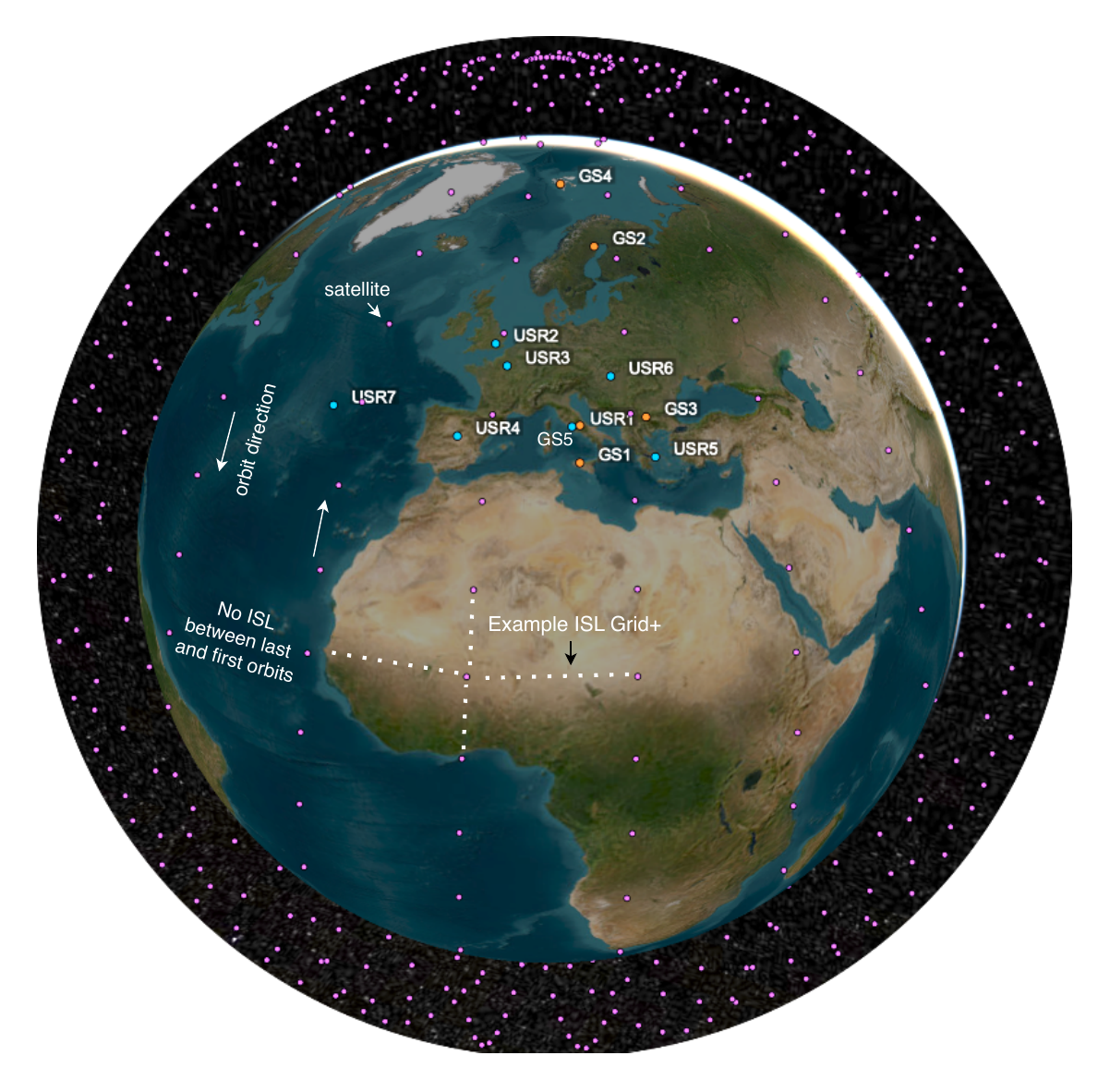}
    \caption{MATLAB visualization of emulated OneWeb-like system with 588 satellites (purple dots), 7 users terminal (USR, light blue dots) and 5 gateways stations (GS, orange dots).}
    \label{fig:snapshot}
\end{figure}

NetSatBench includes a built-in physical-modeling pipeline for scenario generation. The pipeline produces node configurations and periodic epoch files, each representing a \textit{snapshot} of the emulated system. \autoref{fig:generator-pipeline} summarizes its structure.

At its core, the pipeline relies on StarPerf~\cite{mohan2024multifaceted} to compute satellite positions, inter-satellite-link (ISL) topology, and ISL propagation delays for Walker Delta/Star and TLE-based constellations using the Grid+ model~\cite{walker1984satellite}. StarPerf stores these data in an H5 file, which is then processed by a NetSatBench extension tool. This extension augments the H5 dataset with users, gateways, and satellite-to-ground links subject to configurable elevation-angle visibility constraints. It also applies customizable physical-layer models for antenna behavior, bitrate, and packet loss through Python plug-ins. NetSatBench invokes these plug-ins through a standardized interface that provides node positions and visibility information and plug-ins return the corresponding link attributes.

Antenna behavior plug-ins can further restrict connectivity beyond a simple elevation-angle threshold, for example, by accounting for a limited number of antennas at a ground terminal. Bitrate and packet-loss plug-ins can then be applied to the remaining links to characterize communication performance as a function of node geometry rather than as fixed constants.

The resulting extended H5 dataset is consumed by two NetSatBench tools. One generates the \texttt{sat-config.json} and the epoch files describing the emulation scenario (\autoref{fig:control-pipeline}). The other produces MATLAB-based dynamic visualizations of the constellation, a snapshot of which is shown in \autoref{fig:snapshot}.

\subsubsection{Slant-Range-Based Bitrate Model}
\label{sec:bitrate_model}

This section describes the bitrate model for satellite-to-ground links implemented as a physical-layer plug-in and used in the SRv6 evaluation in \autoref{sec:evaluation}. The model is intentionally simple and captures a key property of LEO systems: the achievable bitrate depends on the slant range because both free-space attenuation and atmospheric attenuation vary with propagation distance. Here, the slant range $L$ is the line-of-sight distance between the satellite and the ground terminal. We therefore derive an analytical expression for the bitrate $R_L$ at slant range $L$. The model is parameterized by the zenith bitrate $R_z$, that is, the bitrate at zenith; the zenith signal-to-noise ratio $\mathrm{SNR}_z$; the atmospheric attenuation at zenith $A^{\mathrm{AT}}_{z,\mathrm{dB}}$; and the slant range $L$. At zenith, the slant range reduces to $L=H$, where $H$ denotes the satellite altitude.

In the following, subscripts $L$ and $z$ denote quantities at slant range $L$ and at zenith, respectively. Assuming Shannon-Hartley behavior, the achievable bitrate $R_L$ is proportional to the spectral efficiency $\log_2(1 + \mathrm{SNR}_L)$. By normalizing to the zenith operating point, we obtain
\begin{equation}
\label{eq:rate_shannon}
R_L = R_z \frac{\log_2(1 + \mathrm{SNR}_L)}{\log_2(1 + \mathrm{SNR}_z)} = R_z \frac{\log_2(1 + \mathrm{SNR}_z \cdot \mathcal{A}_L)}{\log_2(1 + \mathrm{SNR}_z)}
\end{equation}
where $\mathcal{A}_L$ is the total link attenuation factor relative to zenith in linear scale. We model it as the product of the free-space term $A^{\mathrm{FS}}$ and the atmospheric-loss term $A^{\mathrm{AT}}$:

\begin{equation}
\label{eq:A}
\mathcal{A}_L = \frac{\mathrm{SNR}_L}{\mathrm{SNR}_z} = \frac{A^{\mathrm{FS}}_z}{A^{\mathrm{FS}}_L} \cdot \frac{A^{\mathrm{AT}}_z}{A^{\mathrm{AT}}_L}
\end{equation}

The free-space term scales with the square of the propagation distance:
\begin{equation}
\label{eq:fs}
\frac{A^{\mathrm{FS}}_z}{A^{\mathrm{FS}}_L} = \left(\frac{H}{L}\right)^2
\end{equation}

The atmospheric attenuation follows the Lambert-Beer law, according to which the received power, in linear scale, decays exponentially with the propagation distance through the absorbing medium. In dB scale, this gives
\begin{equation}
\label{eq:at0}
A^{\mathrm{AT}}_{L,\mathrm{dB}}= 10 \mathrm{log}_{10}(\mathrm{e}) \gamma L
\end{equation}
where the absorption coefficient $\gamma$ depends on atmospheric conditions and carrier frequency. Let $A^{\mathrm{AT}}_{z,\mathrm{dB}}$ denote the atmospheric attenuation at zenith. Using \autoref{eq:at0}, the atmospheric attenuation at slant range $L$ can be written as
\begin{equation}
\label{eq:at}
A^{\mathrm{AT}}_{L,\mathrm{dB}}
=
A^{\mathrm{AT}}_{z,\mathrm{dB}} \frac{L}{H}
\end{equation}

By expressing \autoref{eq:at} in linear scale and combining it with \autoref{eq:fs}, \autoref{eq:A} becomes as follows and can be used in \autoref{eq:rate_shannon}
\begin{equation}
\label{eq:at1}
\mathcal{A}_L = \left(\frac{H}{L}\right)^2 10^{-\frac{A^{\mathrm{AT}}_{z,\mathrm{dB}}}{10}\left(\frac{L}{H}-1\right)}
\end{equation}

\subsection{Routing Framework}
\label{sec:routing}

NetSatBench supports Layer~3 routing through Python plug-ins invoked by the \textit{sat-agent} when topology-relevant events occur, as shown in \autoref{fig:arch}. This design separates link orchestration from routing logic: the emulator manages link creation and removal, while routing plug-ins react to those events by configuring IP forwarding. This enables different routing strategies without modifying the emulator core.

NetSatBench currently provides three built-in routing plug-ins: an IPv4 IS-IS plug-in, an IPv6 IS-IS plug-in, and a \textit{local-routes IPv6 plug-in}. The two IS-IS plug-ins dynamically configure the Linux
routing framework FRR inside the node container. The local-routes IPv6 plug-in instead installs \texttt{/128} routes only to directly connected neighbors through link-local addresses and keeps IPv6 neighbor resolution warm through periodic pings, thereby avoiding RTT spikes caused by the Neighbor Discovery Protocol.

\subsubsection{Epoch-Driven Route Injection}

NetSatBench also allows for route control through epoch files. Epoch files can embed Linux routing commands under the \texttt{run} JSON key, as shown for \texttt{sat1} in \autoref{l:epoch-file-example}. This mechanism enables the evaluation of routing ideas without implementing them inside routing frameworks such as FRR. A routing strategy under test can observe link-state information from Etcd or from the epoch files, compute the required route updates, and inject the corresponding commands into new or updated epoch files, for example, as \texttt{ip route replacement} tasks.

\subsubsection{Idealized Time-Dependent Pre-computed Routing}
\label{sec:oracle}
Using this route-injection mechanism, the NetSatBench toolkit provides an idealized time-dependent routing utility called \textit{oracle routing}. Oracle routing pre-computes route changes before the emulation starts and injects the corresponding update events into the relevant epoch files. Routes are computed with Dijkstra's shortest-path algorithm using either hop count or propagation delay as the metric. Route computation can also be restricted to selected classes of source-destination pairs, for example, satellite-to-satellite or satellite-to-gateway paths.

Oracle routing also implements a \textit{drain-before-break} policy. A few seconds before a link is removed during the emulation, the link is removed from the graph used by Dijkstra's algorithm. The resulting routes, therefore, avoid that link, so no new traffic is forwarded over it. This allows packets already queued in the link buffer to drain before the link is torn down, thereby reducing unnecessary packet loss during handover.

\section{SRv6-based LEO architecture}
\label{sec:usecase}
In this section, we present an SRv6-based LEO architecture that is implemented with real software rather than simulation code and evaluated in NetSatBench, thereby demonstrating both its feasibility and its effectiveness. The design follows the tunneling paradigm used in terrestrial mobile networks, where user traffic is encapsulated in tunnels between the Internet gateway and the user terminal. This avoids maintaining per-user routes in transit nodes, which scales poorly with mobility and large numbers of users. Instead, user-specific state is confined to the tunnel endpoints, while the control plane dynamically updates the tunnel paths.

This rationale naturally extends to LEO systems. If the constellation is viewed as the network core, satellites act as transit nodes, while the access segment changes with the serving satellite. We therefore adopt a tunnel-based design in which the constellation maintains only infrastructure routes and user reachability is ensured by reconfiguring tunnel paths from ground gateway. We refer to these gateway-constellation-user tunnels as \textit{Satellite PDU sessions} because they resemble 5G PDU sessions while extending them to scenarios in which both the user-side and gateway-side access segments may vary.

We adopt Segment Routing over IPv6 (SRv6) as the tunneling mechanism. SRv6 supports IP-in-IP tunneling and controls the tunnel path through an ordered list of segment identifiers (\textit{SIDs}) carried in the IPv6 header of the outer packet, each identifying a landmark node, thereby enabling loose source-controlled routing without dedicated signaling~\cite{ventre2020segment,liu2019segmentlb,he2025srv6rl,zhang2025source}. In our design, SRv6 tunnel paths are entirely controlled by the gateway nodes, allowing traffic-engineering policies to be updated at the edge without modifying the satellite payload.

\begin{figure*}
    \centering
    \includegraphics[scale=0.8]{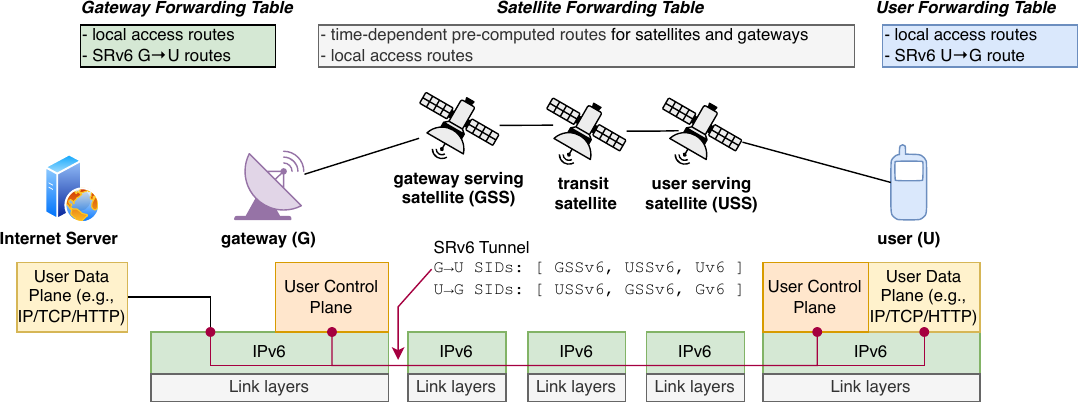}
    \caption{SRv6-based LEO architecture.}
    \label{fig:routing1}
\end{figure*}

\subsection{Protocol Architecture}
\autoref{fig:routing1} shows the considered architecture. The design focuses on user-to-Internet communication and includes only the components needed to assess the feasibility of SRv6 Satellite PDU Sessions in LEO scenarios with real software and procedures. It is organized into the following layers:
\begin{itemize}
\item \textbf{User data-plane:} end-to-end protocols (e.g., IP/TCP/HTTP).
\item \textbf{User control-plane:} procedures for the configuration and updates of the SRv6 tunnels.
\item \textbf{IPv6 layer:} IPv6 forwarding with SRv6 support.
\item \textbf{Link layers:} link and physical-layer protocols supporting link-level communications.
\end{itemize}
Control-plane functions run exclusively on ground nodes, while satellites support only IPv6 forwarding and link-layer operations.

\subsection{Data-Plane Routing and Forwarding}
\subsubsection*{Hybrid IPv6 Routing}
\label{sec:usecase-routing}
IPv6 forwarding follows a hybrid design because its route classes are managed by different mechanisms, as illustrated in \autoref{fig:routing1}. In particular, it relies on the following route classes:
\begin{itemize}
    \item \textbf{Local access routes:} point-to-point routes between satellites and their currently connected ground nodes, managed by a local routing mechanism.
    \item \textbf{Infrastructure routes:} time-dependent precomputed satellite-to-gateway and satellite-to-satellite routes installed on every satellite.
    \item \textbf{SRv6 routes:} gateway-to-user and user-to-gateway routes specifying the SID lists of the corresponding SRv6 tunnels, managed by the control-plane and installed on the gateway and the user, respectively.
\end{itemize}

\subsubsection*{SRv6 Tunneling}
In the user-to-Internet direction, user-plane packets are first steered to the user-serving satellite (USS), then, if distinct, to the gateway-serving satellite (GSS) possibly through transit satellites, and finally to the gateway (G). To realize this forwarding path, the user terminal (U) encapsulates user-plane packets into outer IPv6 \textit{transit} packets with SID list \texttt{[USSv6, GSSv6, Gv6]}, where \texttt{Xv6} denotes the IPv6 address of node \texttt{X}. 
The gateway, which owns the last SID \texttt{Gv6}, decapsulates the outer transit packet and forwards the inner user-plane packet to the Internet according to standard IP processing. In the reverse direction, the gateway encapsulates user-plane packets into outer IPv6 transit packets with SID list \texttt{[GSSv6, USSv6, Uv6]}, steering traffic through the GSS, USS and the user terminal.

The SRv6 overhead is limited to three IPv6 addresses, or two when the GSS coincides with the USS, regardless of how many transit nodes are traversed between consecutive SIDs, thereby supporting scalability in large constellations. Additional intermediate SIDs can be inserted when needed, for example, to enforce a specific detour within the constellation.

\subsection{Control-Plane Procedures}
\label{sec:control-plane}
\subsubsection*{Control Messages}
\begin{figure}
    \includegraphics[scale=0.8]{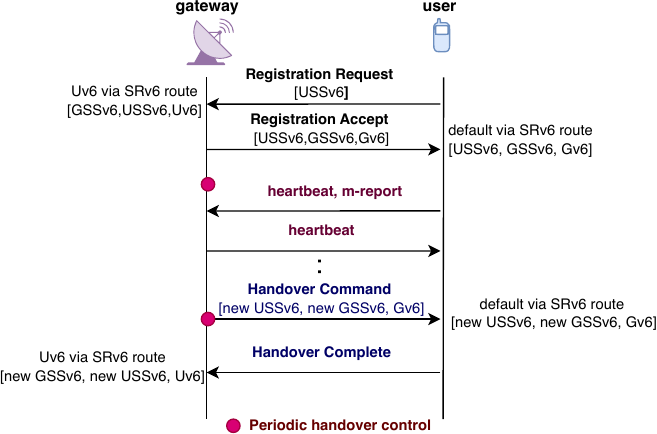}
    \caption{Control-plane message exchange.}
    \label{fig:routing2}
\end{figure}

\autoref{fig:routing2} summarizes the control-plane procedures used to configure and update SRv6 tunnels. A user terminal connects to a single satellite at a time and periodically measures or estimates link-quality parameters, for example, delay, SNR, bitrate, and visibility duration. Gateways can connect to multiple satellites and perform similar measurements.

At initialization, the user selects a gateway, for example the closest one, and a serving satellite (USS) according to a local policy, then sends a \texttt{Registration Request} via the USS, including the IPv6 address of that USS in the payload. Upon reception, the gateway selects the gateway-serving satellite (GSS), installs the forward SRv6 route \texttt{[GSSv6, USSv6, Uv6]}, and replies with a \texttt{Registration Accept} containing the reverse SID list \texttt{[USSv6, GSSv6, Gv6]}. The user installs this reverse path as the default route.

During normal operation, the user and the gateway periodically exchange \texttt{heartbeat} messages, while the user sends link-measurement reports (\texttt{m-report}). When a handover is required, the gateway computes a new SID list, sends a \texttt{Handover Command} over the old SRv6 tunnel, and temporarily pauses traffic to reduce packet loss during the user-side switch. Upon receiving the command, the user switches USS if needed, updates its default route with the new SID list, replies with \texttt{Handover Complete} over the new tunnel, and temporarily pauses traffic for $\mathrm{T}_{\mathrm{ho}}$ to reduce possible packet reordering. The gateway then updates its forward SRv6 tunnel accordingly and resumes traffic. If \texttt{Handover Complete} is not received within $\mathrm{T}_{\mathrm{ho}}$ seconds, the gateway cancels the handover procedure and restores traffic. Heartbeat failures trigger a new registration procedure. 

\subsubsection*{End-to-End Handover Strategy}
\label{sec:ho}
For each user, the gateway periodically checks whether a handover is needed by evaluating the time since the last handover and the remaining lifetime of the current USS and GSS. Handover is triggered if the remaining lifetime falls below $\mathrm{T}_\mathrm{lt}$ or the elapsed time exceeds $\mathrm{T}_\mathrm{el}$; otherwise, the current configuration is kept.

The literature mainly considers \textit{local} handover strategies, which optimize a local objective, such as selecting the access satellite with the longest remaining visibility time or the shortest distance~\cite{lee2023delhandover,wang2023seamless,eydian2025bipartite,voicu2024handover}. Unlike terrestrial networks, satellite handover involves two access links rather than one. This motivates the exploration of \textit{end-to-end} strategies that jointly select the gateway-serving and the user-serving satellites, enabling new objectives such as minimizing the number of traversed ISLs.

Accordingly, we devise an exemplary \textit{end-to-end} strategy in which candidate GSS--USS pairs are progressively filtered through a sequence of policy rules. The last filter is designed to return at most one candidate pair. If no candidate remains after filtering, the current GSS--USS pair is retained and no handover is performed. The implemented policy filters are summarized in \autoref{tab:e2e-filters}.

\begin{table}[t]
\centering
\caption{End-to-End Policy Filters}
\label{tab:e2e-filters}
\begin{tabular}{|p{0.26\linewidth}|p{0.64\linewidth}|}
\hline
\textbf{Filter} & \textbf{Effect} \\ \hline
\textbf{Minimum lifetime} & Discard any GSS--USS pair for which the remaining visibility time of either the GSS or the USS is smaller than $\mathrm{T}_\mathrm{lt}$. \\ \hline
\textbf{Minimum orbit hops} & Retain only the GSS--USS satellite pairs with the minimum ISL hop count among all candidates. \\ \hline
\textbf{Max-min visibility} & Define the visibility of a GSS--USS pair as the minimum of the remaining visibility times of the GSS and the USS. Retain only the GSS--USS pairs with the largest such value among all candidates. \\ \hline
\textbf{Min access delay} & Define the access delay of a GSS--USS pair as the sum of the access delays of the GSS and the USS. Retain only the GSS--USS pairs with the minimum access delay among all candidates. \\ \hline
\textbf{Max-min access rate} & Define the access rate of a GSS--USS pair as the minimum of the link bitrates of the GSS and the USS. Retain only the GSS--USS pairs with the maximum access rate among all candidates. \\ \hline
\end{tabular}
\end{table}

\section{Performance Evaluation}
\label{sec:evaluation}

\subsection{Experimental Setup}

This evaluation illustrates the kind of study that NetSatBench enables through the SRv6 case study, while also reporting the main deployment overheads of the emulation platform. We used a cluster of 10 OpenStack virtual machines as worker nodes. Overall, the cluster provided 72 virtual CPUs (Intel Xeon processors at 3\,GHz) and 72\,GB of memory.

We emulated a OneWeb-like Walker-star constellation with 588 satellites, 5 ground stations, and 7 users, located as shown in \autoref{fig:snapshot}. \autoref{tab:scenario} summarizes the main parameters of the scenario. This setup is intentionally simplified to illustrate NetSatBench's ability to compare alternative handover decisions for the considered SRv6 use case. The ISL topology follows the Grid+ model, in which each satellite maintains links with its immediate neighbors. In the considered Walker-star constellation, the first and last orbital planes move in opposite directions; therefore, no ISLs are established between these two planes.

The system state was updated every 5\,s through epoch files. To limit update overhead, we quantized link delay to 1\,ms and bitrate to 1\,Mbit/s. Accordingly, a link update was included in an epoch file only when the delay changed by more than 1\,ms or the bitrate changed by more than 1\,Mbit/s.

For infrastructure routes, we used the \textit{oracle routing} tool introduced in \autoref{sec:oracle}, with hop count as the metric and the drain-before-break policy enabled. Local IPv6 access routes were handled by the local-routes IPv6 plug-in. Custom gateway and user agents, deployed on the corresponding containers, managed the SRv6 routes and implemented the control-plane procedures described in \autoref{sec:control-plane}.

Deploying all emulated nodes required 1\,min and 32\,s. The initial creation of 1266 links, including insertion into the Etcd database and propagation of link events to the sat-agents, required about 1.12\,s. Each sat-agent then spent about 80\,ms configuring its VXLAN links in parallel. During the experiments, each node container consumed approximately 1.7\% of one CPU core and 30\,MB of memory.

\begin{table}[t]
\centering
\caption{Evaluated Scenario Configuration}
\label{tab:scenario}
\begin{tabular}{|l|l|}
\hline
\textbf{Category} & \textbf{Configuration} \\ \hline

\multicolumn{2}{|c|}{\textit{Constellation}} \\ \hline
Altitude -- Inclination & 1200 km -- $87.9^\circ$ \\
N. of orbits -- Satellites per orbit & 12 -- 49 \\
ISL Topology -- Rate -- Loss & Grid+ -- 400 Mbit/s -- 0\% \\ \hline

\multicolumn{2}{|c|}{\textit{Ground section}} \\ \hline
Visibility min. elevation angle & $25^\circ$ \\
Rate model & Slant-Range-Based \\
Zenith atmospheric loss & 0.5 dB \\
Gateway links zenith rate -- SNR & 200 Mbit/s -- 30 dB \\
User links zenith rate -- SNR & 50 Mbit/s -- 12 dB \\ \hline

\multicolumn{2}{|c|}{\textit{Handover Parameters}} \\ \hline
Min. remaining lifetime $\mathrm{T}_\mathrm{lt}$ & 60\,s \\
Elapsed time threshold $\mathrm{T}_\mathrm{el}$ & 15\,s \\
Periodic control interval & 5\,s \\
Traffic pause $\mathrm{T}_\mathrm{ho}$ & 80\,ms \\  
Shortest-min-access-delay & Policy filter sequence: 1,2,4 \\
Shortest-max-min-access-rate & Policy filter sequence: 1,2,5 \\
Shortest-max-min-visibility & Policy filter sequence: 1,2,3 \\\hline
\end{tabular}
\end{table}

\subsection{RTT analysis}
We analyze the RTT measured by users while sending \texttt{ping} probes to their gateway every 10\,ms. We first consider the widely studied \textit{min-access-delay} local handover strategy \cite{voicu2024handover}. Under this policy, the gateway and the user independently trigger a handover when they can attach to a satellite with a lower access delay than the currently selected one.

\autoref{fig:intro-rtts} in \autoref{sec:intro} reports the related RTT values measured for users~1 and~7, both connected to gateway~5. RTT spikes occur during handovers because traffic is temporarily paused\footnote{We emulate a traffic pause by temporarily reducing the user bitrate so that the first queued packet has a transmission time of 80 ms. This approach is not always exact, but it avoids the packet reordering that would arise if the pause were emulated by adding delay with Linux \texttt{tc netem}.}. Between handovers, the RTT still exhibits smaller variations. Some of them reflect actual orbital dynamics, whereas others are minor emulation artifacts caused by the snapshot-based and quantized system updates of links' delay and bitrate.

The large difference in RTT between users~1 and~7 highlights a severe limitation of local handover strategies. User~1 and gateway~5 independently select the same access satellite, allowing for bent-pipe forwarding, or two satellites separated by at most one ISL; as a result, the RTT evolves in steps of approximately 5--10\,ms and remains below about 40\,ms. In contrast, as shown in \autoref{fig:snapshot}, user~7 is located at sea and selects a satellite in the descending orbital plane on its left because that satellite offers the lowest local access delay, although this is not immediately apparent in \autoref{fig:snapshot} because of the viewing perspective. Gateway~5 instead selects a satellite in a nearby ascending orbital plane. Since no ISLs are established between the first and last orbital planes because of their opposite directions, the resulting end-to-end path must traverse a long sequence of ISLs around the globe, increasing the RTT from tens to hundreds of milliseconds. In principle, user~7 could have selected a satellite in an ascending orbital plane and avoided this detour. However, the local policy optimizes only the user-side access link and does not account for the choice made at the opposite endpoint.

\begin{figure}
    \centering
    \includegraphics[width=0.8\linewidth]{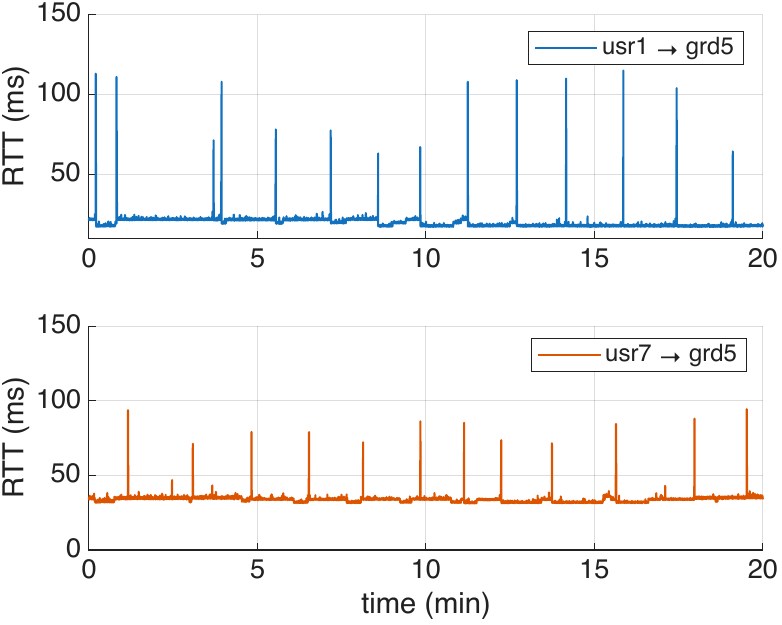}
    \caption{RTT measurements between users 1,7 and the gateway 5 in a OneWeb-like scenario. End-to-end shortest-min-access-delay handover strategy}
    \label{fig:coupled-min-delay}
\end{figure}

\autoref{fig:coupled-min-delay} shows the RTT under the end-to-end handover strategy denoted as \textit{shortest-min-access-delay} in \autoref{tab:scenario}, where the term \textit{shortest} indicates the presence of the minimum-orbit-hops policy. Compared with \autoref{fig:intro-rtts}, the RTT of user~7 drops markedly to about 35\,ms because the minimum-orbit-hops filter encourages the gateway and the user to select, whenever possible, either the same satellite or a pair of satellites connected by only a few ISLs. User~1 also experiences a lower RTT during some intervals, especially before 10\,mins, for the same reason. Unlike the end-to-end strategy, a local policy cannot identify the available bent-pipe configuration that is most effective in reducing delay and ISL hops.

\begin{figure}
    \centering
    \includegraphics[width=0.8\linewidth]{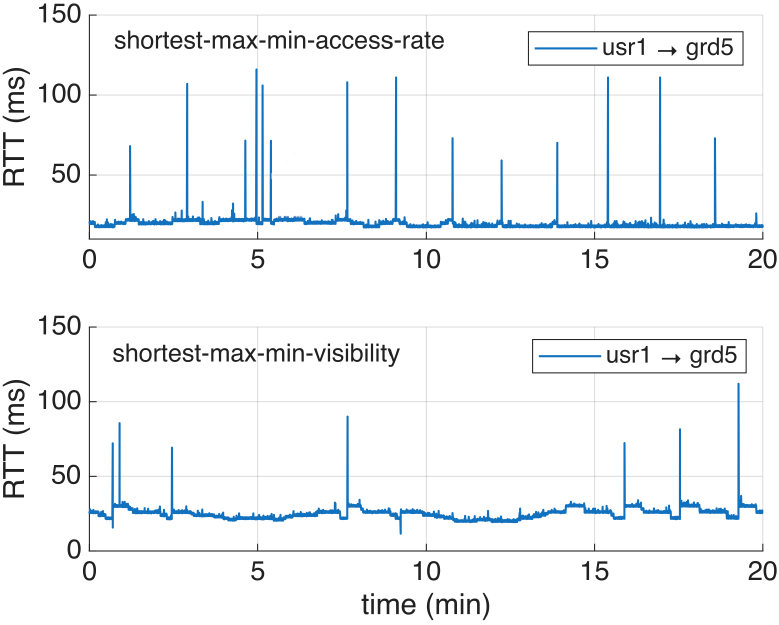}
    \caption{RTT measurements between a user and the gateway in a OneWeb-like scenario for shortest-max-min-rate and shortest-max-min-visibility end-to-end handover strategies}
    \label{fig:coupled-rate-visibility}
\end{figure}

\autoref{fig:coupled-rate-visibility} reports the RTT of user~1 under the end-to-end handover strategies \textit{shortest-max-min-access-rate} and \textit{shortest-max-min-visibility}. Compared with \autoref{fig:coupled-min-delay}, the rate-optimized strategy does not exhibit a relevant RTT difference because, in our slant-range-based model, rate and delay are directly correlated. By contrast, the visibility-oriented strategy aims to reduce the number of handovers by selecting, and keeping, whenever possible, the satellite with the longest visibility period. Consequently, the number of handover events, marked by RTT spikes, is significantly lower, although the different optimization objective cause slight increases in RTT.

\subsection{TCP analysis}
\begin{figure}
    \centering
    \includegraphics[width=0.8\linewidth]{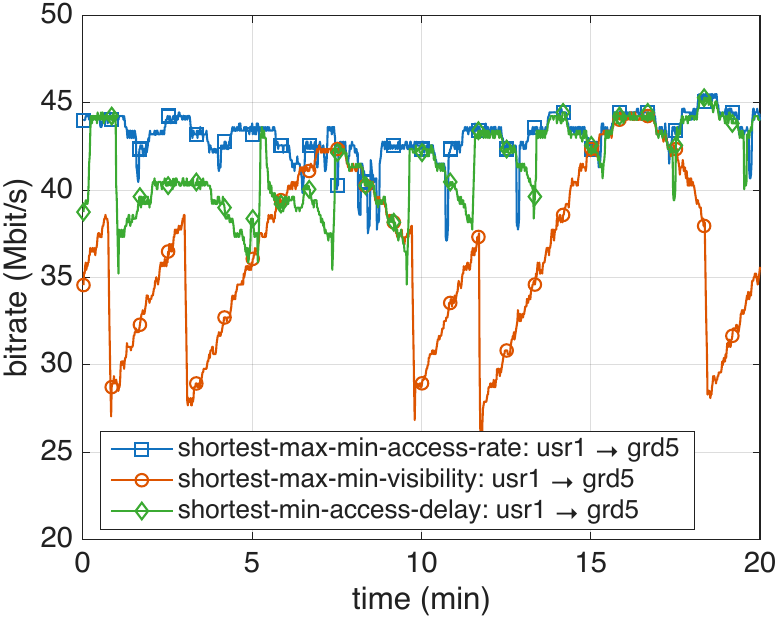}
    \caption{TCP-cubic bitrate between user~1 and the gateway 5 in a OneWeb-like scenario for different end-to-end handover strategies}
    \label{fig:cubic-comparison}
\end{figure}

\autoref{fig:cubic-comparison} compares the three end-to-end handover strategies in terms of TCP bitrate, measured with the \texttt{iperf3} tool using Cubic congestion control between user~1 and gateway~5. The rate-optimized strategy achieves higher throughput for most of the experiment. All strategies exhibit an oscillatory throughput pattern, which is induced by the slant-range-based rate model. When an access satellite is selected while rising above the horizon, its slant range decreases, and the achievable bitrate increases. After the satellite reaches its minimum slant range, it starts moving toward the horizon, causing the slant range to increase and the throughput to decrease. This effect is more pronounced for the visibility-optimized strategy, which tends to keep the same access satellite for as long as possible.
Taking into account both RTT and throughput, we identify the shortest-max-min-access-rate strategy as a valuable tradeoff among the evaluated strategies.

\begin{figure}
    \centering
    \includegraphics[width=0.9\linewidth]{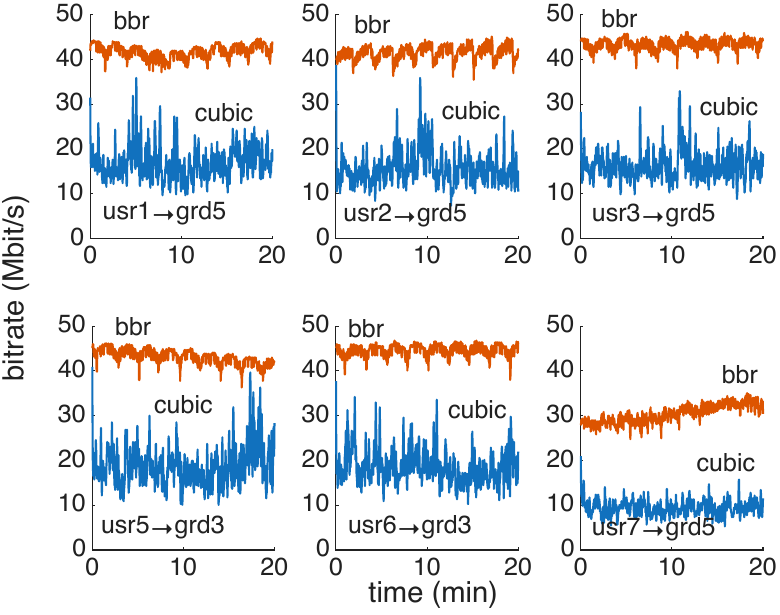}
    \caption{TCP-cubic (blue) and TCP-bbr (red) bitrate between users and the gateways in a OneWeb-like scenario for shortest-max-min-access-rate end-to-end handover strategies with random packet loss at 0.1\% on access links}
    \label{fig:bbr-vs-cubic}
\end{figure}

We conclude the evaluation by analyzing the impact of TCP congestion control. Previous studies have observed that loss-, timeout-, or reordering-driven congestion control algorithms, such as Cubic, can underperform in LEO scenarios because these events may be caused by temporary link impairments or handovers rather than persistent congestion. Rate-estimation-based algorithms, such as Google's BBR, can therefore be more robust in this setting~\cite{barbosa2023comparative,garcia2025tcp}. \autoref{fig:bbr-vs-cubic} illustrates this effect by comparing the standard Linux implementations of Cubic and BBR under the rate-optimized handover strategy, with a fixed random packet loss of 0.1\% applied to all access links. In this lossy scenario, BBR shows stronger resilience to random packet loss than Cubic. When we repeat the experiment without synthetic packet loss, the two congestion control algorithms do not exhibit marked performance differences. 
 
\section{Related Works}
\label{sec:related}
\subsection{Container-based LEO Emulators}
Container-based emulators occupy a middle ground between packet-level simulators and VM-based testbeds \cite{pfandzelter2022celestial}. Unlike simulators, they can execute real software stacks and application binaries. Unlike VM-based platforms, they typically scale to larger constellations because containers have a smaller resource footprint, at the cost of weaker isolation and limited support for per-node kernel modifications unless the worker-node kernel is modified.

LeoEM \cite{cao2023satcp} adopts a Mininet-based design in which LEO nodes are emulated with Docker containers. This approach is useful for functional validation, but Mininet-based systems typically rely on a single host kernel and therefore scale poorly once the emulated constellation grows.

StarryNet~\cite{lai2023starrynet} adopts a container-based architecture in which satellites, gateways, and user terminals are instantiated as Docker containers running real networking stacks. The platform models the links through constant-rate abstractions and supports OSPF as its main routing protocol. Addressing is configured automatically in IPv4, with limited customization, and experiment execution mainly targets predefined measurements such as \texttt{ping} and \texttt{iperf}. The system relies on a centralized orchestrator to update link states, and each emulated link between connected nodes is realized through a Linux bridge and a pair of \texttt{veth} interfaces connecting the node containers to that bridge. Although the paper discusses distributed deployment using VLANs, this functionality is not publicly available in the current implementation\footnote{We refer to the state of the GitHub repository as of April 21, 2026.}.

OpenSN~\cite{lu2025opensn} introduces eBPF-based mechanisms for efficient emulated-link creation and reconfiguration. In its high-performance mode, links are realized through MAC-address translation performed by an eBPF module, without per-link VLAN/VXLAN overlays or Linux bridges, thereby reducing setup time and network overhead. OpenSN allows users to define custom node roles through custom Docker images, thus supporting arbitrary software inside the emulated nodes. From a topology-modeling perspective, it relies on TLE data to compute satellite positions and on a centralized Topology Configurator to update connectivity over time, while link characteristics are specified through fixed parameters such as bitrate and packet loss. The control plane distributes topology updates through an Etcd key-value store, and local agents on worker nodes translate these updates into eBPF networking primitives.

Compared with these systems, NetSatBench emphasizes a clearer separation between the core emulation infrastructure and experiment-specific logic. In particular, topology evolution is driven by declarative epoch files, while physical-layer and routing behavior can be customized through external tools and plug-in interfaces rather than by modifying the emulator internals. More generally, whereas StarryNet and OpenSN primarily expose programmatic interfaces for defining topology evolution and experiment actions, NetSatBench adopts a higher-level declarative workflow in which researchers describe scenarios and events in JSON files and execute the emulation through a command-line interface. Besides, a richer set of built-in routing and physical-layer models is supported along with IPv6. From an operational perspective, NetSatBench also provides a \texttt{key:value} node-selection mechanism that simplifies large-scale experiment management by enabling targeted configuration, file transfer, and command execution. \autoref{tab:comparison} summarizes the main differences between NetSatBench and representative distributed container-based emulators.

\begin{table*}[t]
\centering
\caption{Comparison of LEO network container-based distributed emulation platforms}
\label{tab:comparison}
\begin{tabular}{l|p{0.2\linewidth}|p{0.2\linewidth}|p{0.3\linewidth}}
\hline
\textbf{Feature} & \textbf{StarryNet} & \textbf{OpenSN} & \textbf{NetSatBench} \\
\hline
Link creation speed & Moderate; per-link \texttt{veth}, VLAN, and bridge setup & Very high; per-link eBPF configuration& High; per-link VXLAN setup\\
\hline
Scalability & Moderate; centralized control, closed-source VLAN support & High; worker-level distributed control & High; container-level distributed control \\
\hline
Topology control & Integrated; customization requires emulator code changes & Integrated; customization requires emulator code changes & Decoupled via epoch files generated or configured by external tools \\
\hline
Experiment workflow & Primarily programmatic & Primarily programmatic & Declarative JSON + CLI workflow \\
\hline

Built-in topology support & Walker-based constellations & TLE-based constellations & Extended StarPerf support for Walker-based and TLE-based constellations \\
\hline
Link control & Integrated; customization requires emulator code changes & Moderately integrated; customization requires emulator code changes or replacement of the Topology Configurator & Plug-in extensibility; no emulator code changes required, and the plug-in receives node physical-layer configuration through a standard interface and simply returns link parameters \\
\hline
Built-in link control & ISL Grid+, elevation-angle visibility, fixed link rate and loss, distance-based delay & ISL Grid+, elevation-angle visibility, fixed link rate and loss, distance-based delay & ISL Grid+, elevation-angle visibility, make-before-break antenna management, fixed or slant-range-based link rate, fixed loss, distance-based delay \\
\hline
IP support & IPv4 with built-in automatic addressing & IPv4 with built-in automatic addressing & IPv4 and IPv6 with highly configurable multi-subnet automatic addressing through \texttt{key:value} matching of node properties\\
\hline
Routing control & Integrated; customization requires emulator code changes & Plug-in extensibility; routing modules inside the container monitor a node-status file updated by the emulator & Plug-in extensibility; routing modules inside the container implement an event-driven interface invoked on link events by the emulator\\
\hline
Built-in routing & OSPFv2 for IPv4 & OSPFv2 for IPv4 & IS-IS for IPv4, IS-IS for IPv6, IPv6 link-local forwarding, and precomputed IPv4/IPv6 routing \\
\hline
Custom software onboarding & Requires Docker image customization & Requires Docker image customization & \texttt{key:value} matching software copy to running nodes; no Docker image rebuild required \\
\hline
\end{tabular}
\end{table*}

\subsection{Segment Routing in LEO Satellite Networks}
In the LEO context, Zhang et al.~\cite{zhang2024link, zhang2025source} motivate segment-routing (SR) as a mechanism to avoid untrusted regions and to differentiate forwarding paths according to QoS requirements. To reduce header overhead over long multi-hop satellite paths, they further propose a Bloom-filter-based encoding of the segment list. Chen et al.~\cite{chen2023fast} partition satellites into computing and forwarding roles so that backup SR paths can be pre-computed on orbit, and only the affected segments need to be updated after failures, thereby reducing dependence on ground control-plane reaction time. Mao et al.~\cite{mao2025intent} propose an intent-driven SR framework with polar-region link handling, dynamic load balancing and multi-path recovery, evaluated in ns-3 simulator over Iridium and OneWeb scenarios.

Unlike these studies, we investigate IPv6 SR as a practical mechanism for realizing satellite PDU sessions together with the associated control-plane procedures. NetSatBench allowed us to evaluate this design with real software under time-varying LEO dynamics.

\section{Conclusions}
\label{sec:conclusion}

NetSatBench shows that large-scale LEO constellations can be emulated with real non-simulated software while preserving the flexibility needed to study time-varying topology, routing, and application behavior. Its declarative workflow and plug-in architecture make it possible to evaluate networking solutions and in-orbit applications without modifying the emulator core.

We used NetSatBench to assess an SRv6-based LEO architecture for end-to-end user--gateway communication. In particular, the plug-in-based physical-modeling support allowed us to incorporate a slant-range-based bitrate model, revealing differences between handover strategies that would be hidden by simpler constant-rate assumptions. The routing and task-injection mechanisms also enabled the evaluation of a hybrid routing framework combining precomputed infrastructure routes with dynamic IPv6 edge behavior.

The case study highlights the importance of end-to-end handover strategies that jointly consider satellites serving the gateway and the user, rather than relying on local decisions only. The proposed SRv6 architecture is still preliminary, but NetSatBench provides a practical foundation for further experimental work.

\bibliographystyle{IEEEtran}
\bibliography{sample}

\end{document}